\documentclass[useAMS,usenatbib]{mn2e}
\usepackage{amsfonts}
\usepackage{graphicx,float}
\usepackage[ansinew]{inputenc}
\usepackage{datetime}
\usepackage{titlesec}
\usepackage{subfig}
\title[The $R_{\mathrm{h}}=ct$ universe and quintessence]{The $R_{\mathrm{h}}=ct$ universe and quintessence}
\author[J. Sultana]{J. Sultana$^{1}$\thanks{E-mail:
joseph.sultana@um.edu.mt}\\
$^{1}$Department of Mathematics, University of Malta, Msida, MSD2080 Malta}
\begin{document}

\date{}

\pagerange{\pageref{firstpage}--\pageref{lastpage}} \pubyear{2015}

\maketitle

\label{firstpage}

\begin{abstract}
Over the last few years the $R_{\mathrm{h}}=ct$ universe has received a lot of attention, particularly when observational evidence seems to favor this over the standard $\Lambda$ cold dark matter ($\Lambda CDM$) universe. Like the $\Lambda CDM$, the $R_{\mathrm{h}}=ct$ universe is based on a Friedmann-Robertson-Walker (FRW) cosmology where the total energy density $\rho$ and pressure $p$ of the cosmic fluid contain a dark energy component besides the usual (dark and baryonic) matter and radiation components. However unlike the $\Lambda CDM$ this model has the simple equation of state (EOS) $\rho + 3p = 0$, i.e., its total active gravitational mass vanishes, which would therefore exclude a cosmological constant as the source of its dark energy component. Faced with this issue, in this paper we examine various possible sources for the dark energy component of the $R_{\mathrm{h}}=ct$ universe and show that quintessence which has been used in other various dynamical dark energy models could also be a possible source in this case.
\end{abstract}


\section{Introduction}
One of the first and simplest proposed Friedmann-Robertson-Walker (FRW) cosmological model is the $\Lambda$ cold dark matter ($\Lambda$CDM) universe, which involves Einstein's cosmological constant $\Lambda$. This standard model of cosmology, which is also referred to as the concordance model, assumes that the total energy density $\rho$ of the universe is made up of three components, namely matter $\rho_{m}$ (baryonic and dark matter), radiation $\rho_{r}$, and dark energy or vacuum energy $\rho_{\Lambda}$, which produces the necessary gravitational repulsion. In this model, dark energy which has an equation of state (EOS) $\omega_{\Lambda} = p_{\Lambda}/\rho_{\Lambda} = -1$, is a property of the space itself and its density $\rho_{\Lambda} = -p_{\Lambda} = \Lambda c^4/8\pi G$ is constant, such that as the universe expands the constant vacuum energy density will eventually exceed the matter density of the universe which is ever decreasing. The spatially flat $\Lambda$CDM model dominated by vacuum energy with $\Omega_{\Lambda} \sim 0.70$, with the rest of the energy density being in the form of nonrelativistic cold dark matter with $\Omega_{m} \sim 0.25$ and nonrelativistic baryonic matter with $\Omega_{b} \sim 0.05$, fits observational data reasonably well \citep{riess98,perlmutter99,knop03,riess04}. However the main problem in this model is the huge difference of about $10^{120}$ orders of magnitude between the observed value of the cosmological constant and the one predicted from quantum field theory; known as the cosmological constant problem \citep{weinberg89}. Another issue is the so called coincidence problem which expresses the fact that although in this model the matter and dark energy components scale differently with redshift during the evolution of the universe, both components today have comparable energy densities, and it is unclear why we happen to live in this narrow window of time.\\
Besides these main issues, there are other inherent problems faced by the $\Lambda$CDM, some of which arose as a result of recent observations that are in disagreement with the model's predictions. For example in order to account for the general isotropy of the cosmic microwave background (CMB), the standard model invokes an early period of inflationary expansion \citep{kazanas80,guth81,linde82}. However the latest observations by \textit{Planck} \citep{ade14} indicate that there may be some problems with such an inflationary scenario \citep{ijjas13,guth14}. It was partly due to these issues of the standard $\Lambda$CDM, that during the last decade several alternative dark energy models have been proposed and tested with observations. In these models the dark energy density component $\rho_{de}$ is not constant and in most cases $\omega_{de} = p_{de}/\rho_{de}$ depends on time, redshift, or scale factor. For example in some of these so called dynamical dark energy models, late time inflation is achieved using a variable cosmological term $\Lambda(t)$ \citep{ray11,basilakos15} sometimes taken in conjunction with a time dependent gravitational constant $G(t)$ \citep{ray07,singh13}. Other sources of dark energy include scalar fields such as quintessence \citep{peebles03}, K-essence \citep{armendariz01} and phantom fields \citep{singh03}. An alternative approach to the dark energy problem relies on the modification of Einstein's theory itself such that in these alternative theories of gravity, cosmic acceleration is not provided solely by the matter side $T_{\mu\nu}$ of the field equations, but also by the geometry of spacetime. These theories include the scalar tensor theory with non-minimally coupled scalar fields \citep{bertolami00,barrow97}, $f(R)$ theory \citep{tsujikawa08}, conformal Weyl gravity \citep{mannheim00} and higher dimensional theories such as the Randall-Sundrum (RS) braneworld model \citep{randall99}, and the braneworld model of Dvali-Gabadadze-Porrati (DGP) \citep{dvali00}.\\
Over the last few years considerable interest has been shown in the simple FRW linearly expanding (coasting) model in Einstein's theory with $a(t) \propto t$, $H(z) = H_{0}(1 + z)$. Like the $\Lambda$CDM the total energy density and pressure in this model are expressed in terms of matter, radiation and dark energy components, such that $p = \omega\rho$ with $\rho = \rho_{m} + \rho_{r} + \rho_{de}$ and $p = p_{r} + p_{de}$ (since $p_{m}\approx0$), but it includes the added assumption $\omega = -1/3$, i.e. the cosmic fluid acting as the source has zero active gravitational mass. So this would definitely exclude a cosmological constant as the source of the dark energy component in this case. The model was first discussed by \cite{kolb89} who referred to this zero active mass cosmic fluid as ``K-matter''.  Interest in this model has been revived recently after it was noted \citep{melia03} that in the standard model the radius of he gravitational horizon $R_{h}(t_0)$ (also known as the Hubble radius) is equal to the distance $ct_0$ that light has traveled since the big bang, with $t_0$ being the current age of the universe. In the $\Lambda$CDM this equality is a peculiar coincidence because it just happens at the present time $t_0$. It was then proposed \citep{melia07,melia09,melia12} that this equality may not be a coincidence at all, and should be satisfied at all cosmic time $t$. {This was done by an application of Birkhoff's theorem and its corollary, which for a flat universe allows the identification of the Hubble radius $R_h$ with the gravitational radius $R_h = 2GM/c^2$, given in terms of the Misner-Sharp mass $M = (4\pi/3)R_{h}^{3}(\rho/c^2)$ \citep{misner64}. The added assumption of a zero active gravitational mass $\rho + 3p = 0$ implies \citep{melia12} that $R_h = ct$ or $H = 1/t$ for any cosmic time $t$. This linear model became known as the $R_h = ct$ universe. Unlike the $\Lambda$CDM/$\omega$CDM which contains at least the three parameters $H_0$, $\Omega_m$ and $\omega_{de}$, the $R_h = ct$ model depends only on the sole parameter $H_0$, so that for example the luminosity distance used to fit Type Ia supernova data \citep{melia09} is given by the simple expression $d_{L} = (1+z)R_{h}(t_{0})\ln(1 +z)$. Also while the $\Lambda$CDM would need inflation to circumvent the well-known horizon problem, the $R_h = ct$ universe does not require inflation. One should also point out that the condition $R_h = ct$ is also satisfied by other linear models such as the Milne universe \citep{milne33}, which however has been refuted by observations. Unlike the $R_h = ct$ model discussed here, the Milne universe is empty ($\rho = 0$) and with a negative spatial curvature ($k = -1$). As a result of these properties its luminosity distance is given by $d_{L}^{\textsuperscript{Milne}} = R_{h}(t_0)(1 + z)\sinh[\ln(1+z)]$, and it was shown that this is not consistent with observational data \citep{melia12}.}\\
In the last few years the $R_h=ct$ universe received a lot of attention when it was shown \citep{melia13,wei13,wei14a,wei14b,wei15,melia15a} that it is actually favored over the standard $\Lambda$CDM (and its variant $\omega$CDM with $\omega \neq -1$) by most observational data. This claim has been contested by \citet{bilicki12} and \citet{shafer15} who argued that measurement of $H(z)$ as a function of redshift and the analysis of Type Ia supernovae favored the $\Lambda$CDM over the $R_h = ct$ universe. However this was later contested by \cite{melia15b} who showed that the $R_h=ct$ was still favored  when using model-independent measurements that are not biased towards a specific model. Others (see for example \cite{van-oirschot10,lewis12,mitra14} have also criticized the model itself, particularly the validity of the EOS $\omega = -1/3$ \citep{lewis13}. These and other criticisms have been addressed by \cite{bikwa12,melia12a} (see also \cite{melia15c} and references therein.)\\
As pointed out above the $R_h = ct$ model would still require a dark energy component $\rho_{de}$, albeit not in the form of a cosmological constant. So the obvious question at this point would be: what are the possible sources for this component that together with the matter and radiation components will give the required total EOS, $\omega = -1/3$? The purpose of this paper is to answer this question by discussing the various possible sources of dark energy that are consistent with this EOS. Since the radiation component $\rho_{r}$ at the present time $t_{0}$ is insignificant (at least for the $\Lambda$CDM with which this model has been compared) we assume that the total energy density $\rho = \rho_{de} + \rho_{m}$ and the total pressure $p = p_{de}$ ($p_{m}\approx0$), as is normally done in the other alternative dynamical dark energy models found in the literature. So in the next three sections we examine three possibilities for the source of dark energy in the $R_h = ct$ model, namely a variable cosmological term $\Lambda(t)$, a non-minimally coupled scalar field in Brans-Dicke theory which is equivalent to a variable gravitational constant $G(t)$, and finally quintessence represented by a minimally coupled scalar field $\phi$. We show that although the first two sources are consistent with the model, they are both unphysical, which leaves the third source of quintessence as the viable source of dark energy in the $R_{h} = ct$ universe. Results are then discussed in the Conclusion. Unless otherwise noted we use units such that $G = c = 1$.

\section[]{Varying cosmological term $\Lambda(t)$}

In this case Einstein's field equations take the form
\begin{equation}
R^{ij} - \frac{1}{2}Rg^{ij} = 8\pi\left[T^{ij} - \frac{\Lambda}{8 \pi}g^{ij}\right], \label{efe}
\end{equation}
where $\Lambda(t)$ is a time varying vacuum term representing the dark energy component of the source.  In this case the time dependent vacuum satisfies the same EOS as in the $\Lambda$CDM, i.e., $w_{de} = p_{de}/\rho_{de} = -1$, but unlike the standard model the energy density $\rho_{de} = \Lambda(t)/8\pi$ is a function of the cosmic time. Taking the spatially flat FRW metric
\begin{equation}
ds^2 = -dt^2 + a(t)^2[dr^2 + r^2(d\theta^2 + \sin^2\theta d\phi^2)], \label{frw}
\end{equation}
where $a(t)$ is the scale factor, and the energy momentum tensor of a perfect fluid
\begin{equation}
T_{ij} = (\rho_m + p_m)u_{i}u_{j} + p_m g_{ij}, \label{em-tensor}
\end{equation}
where $u^{i} = [1,0,0,0]$ is the four-velocity vector in comoving coordinates,
the field equations in (\ref{efe}) give
\begin{eqnarray}
\left(\frac{\dot{a}}{a}\right)^2  & = & \frac{8\pi}{3}\rho_{m} + \frac{\Lambda}{3}, \nonumber \\
\frac{\ddot{a}}{a} & = & -\frac{4\pi}{3}(\rho_{m} + 3p_{m}) + \frac{\Lambda}{3}. \label{efeqs}
\end{eqnarray}
These equations can also be combined into
\begin{equation}
\left(\frac{\dot{a}}{a}\right)^2 + \frac{2}{(1 + 3\omega_{m})}\frac{\ddot{a}}{a} = \left(\frac{1+\omega_{m}}{1 + 3\omega_{m}}\right)\Lambda, \label{combined1}
\end{equation}
where $p_{m} = \omega_{m}\rho_{m}$. Using the Hubble parameter $H = \dot{a}/a$ this can also be written as
\begin{equation}
H^2 + \frac{2}{1 + 3\omega_{m}}(\dot{H} + H^2) = \left(\frac{1+\omega_{m}}{1 + 3\omega_{m}}\right)\Lambda. \label{combined2}
\end{equation}
From Bianchi identities we get
\begin{equation}
\dot{\rho}_{m} + 3(1 + \omega_{m})H\rho_{m} + \dot{\rho}_{de} = 0,
\end{equation}
so that in the case of a time dependent cosmological term $\Lambda(t)$, the matter part cannot be conserved separately as in the $\Lambda$CDM where $\rho_{m} \sim a^{-3}$, and so there should be some energy exchange between the two components.\\
For the $R_{h} = ct$ model, $a(t) = t/t_{0}$ where $t_{0}$ is the current age of the universe, so that $H(t) = 1/t$. As usual for the matter part we can take $\omega_m = 0$, so that (\ref{combined2}) gives a time decreasing vacuum term $\Lambda = 1/t^2$. The matter energy density is then obtained from (\ref{efeqs}) and is given by $\rho_{m} = \frac{1}{4\pi t^2}$, which shows that it is not separately conserved. In this case the cosmological parameters $\Omega_{m} = \frac{8\pi}{3H^2}\rho_m = 2/3$ and $\Omega_{de} = \Lambda/3H^2 = 1/3$ are constants. This contrasts with the current value of $\Omega_{m0} \approx 0.27$ which is used in the $\Lambda$CDM to adequately fit the data. Also a similar and more general cosmological model in which the cosmological term $\Lambda = 3\gamma H^2$ and $a(t) = [3(1 - \gamma)H_0 t/2]^{2/3(1 - \gamma)}$ was discussed by \citet{basilakos09} (see also \citet{arcuri94}) and it was shown that for $\gamma \geq 1/3$, cosmic structures cannot be formed via gravitational instability in such models. This therefore suggests that a variable cosmological term (with $\Lambda = 1/t^2$ corresponding to $\gamma = 1/3$) may not be an appropriate explanation for dark energy in the $R_h = ct$ universe.

\section[]{Varying gravitational term $G(t)$}

An example of an alternative gravitational theory to general relativity, in which the gravitational constant $G$ is a function of spacetime is given by Brans-Dicke (BD) theory \citep{brans61}. The original motivation for the introduction of BD-theory was to produce a theory that accommodates Mach's principle \citep{sciama53} which is not completely consistent with general relativity. The theory satisfies Dirac's Large Number Hypothesis (LNH) and the variable gravitational term $G \sim 1/\psi$ is expressed in terms of a scalar field. BD theory has seen a renewed interest due to its association with superstring theories, extra-dimensional theories and cosmological models with inflation or accelerated expansion \citep{callan85,duff95,banerjee01,fabris06}. The action of BD-theory in the so called Jordan frame is given by
\begin{equation}
S^{(\mathrm{BD})} = \frac{1}{16\pi}\int d^4x\sqrt{-g}\left[\psi R - \frac{\omega}{\psi}g^{cd}\nabla_{c}\psi\nabla_{d}\psi - V(\psi)\right] + S^{(\mathrm{M})}, \label{action}
\end{equation}
where
\begin{equation}
S^{(\mathrm{M})} = \int d^4x\sqrt{-g}\mathcal{L}^{\mathrm{M}}
\end{equation}
is the matter action and $\omega$ is the dimensionless Brans-Dicke parameter\footnote{In this section the Brans-Dicke parameter $\omega$ should not be confused with the EOS parameter $\omega$ for the $R_h = ct$ universe.}. BD theory approaches general relativity in the limit $|\omega|\rightarrow \infty$ and $V(\phi)\rightarrow 0$. Moreover solar system constraints from the Cassini mission \citep{bertotti03} imply that $\omega > 4\times10^4$. In our case we take the scalar field potential $V(\psi)$ to be zero. Varying the action with respect to the metric tensor $g_{ab}$ gives
\begin{equation}
G_{ab}  =  \frac{8\pi}{\psi}T_{ab} + \frac{\omega}{\psi^2}\left(\nabla_{a}\psi\nabla_{b}\psi - \frac{1}{2}g_{ab}\nabla^{c}\psi\nabla_{c}\psi\right)
+ \frac{1}{\psi}\left(\nabla_{a}\nabla_{b}\psi - g_{ab}\Box\psi\right), \label{bdeqs}
\end{equation}
where $\Box\psi = g^{ab}\nabla_{a}\nabla_{b}\psi$ and
\begin{equation}
T_{ab} = \frac{-2}{\sqrt{-g}}\frac{\delta}{\delta g^{ab}}\left(\sqrt{-g}\mathcal L^{(\mathrm{M})}\right).
\end{equation}
Varying the action with respect to the scalar field gives
\begin{equation}
\frac{2\omega}{\psi}\Box\psi + R - \frac{\omega}{\psi^2}\nabla^{c}\psi\nabla_{c}\psi = 0. \label{sfeq}
\end{equation}
Taking the trace of (\ref{bdeqs}), yields
\begin{equation}
R = \frac{-8\pi T^{(\mathrm{M})}}{\psi} + \frac{\omega}{\psi^2}\nabla^{c}\psi\nabla_{c}\psi + \frac{3\Box\psi}{\psi},
\end{equation}
such that the scalar field equation in (\ref{sfeq}) becomes
\begin{equation}
\Box\psi = \frac{8\pi}{2\omega + 3}T^{(\mathrm{M})}, \label{fieldeq}
\end{equation}
where $T^{(\mathrm{M})} = T^{\mu}_{\mu}$ is the trace of the matter energy momentum tensor. Multiplying (\ref{bdeqs}) by $\psi$ and taking the covariant derivative leads to
\begin{equation}
8\pi T^{a}_{b;a} = -\frac{1}{2}\left(R - \frac{\omega}{\psi^2}\psi_{,a}\psi^{,a} + \frac{2\omega}{\psi}\Box\psi\right)\psi_{,b},
\end{equation}
such that unlike the previous case (\ref{sfeq}) implies the separate conservation of the matter energy momentum tensor $T^{a}_{b;a} = 0$. \\
For the FRW metric in (\ref{frw}) and the matter energy momentum tensor in (\ref{em-tensor}) the field equations in (\ref{bdeqs}) and (\ref{fieldeq}) lead to the following equations
\begin{eqnarray}
\frac{3\dot{a}^2}{a^2} & = & \frac{8\pi}{\psi}\rho_m + \frac{\omega}{2}\frac{\dot{\psi}^2}{\psi^2} - \frac{3\dot{a}}{a}\frac{\dot{\psi}}{\psi}, \label{eq1} \\
-\frac{2\ddot{a}}{a} - \frac{\dot{a}^2}{a^2} & = & \frac{8\pi}{\psi}p_{m} + \frac{\omega}{2}\frac{\dot{\psi}^2}{\psi^2} + \frac{\ddot{\psi}}{\psi} + \frac{2\dot{a}}{a}\frac{\dot{\psi}}{\psi}, \label{eq2}\\
\frac{\ddot{\psi}}{\psi} + 3\frac{\dot{a}}{a}\frac{\dot{\psi}}{\psi} & = & \frac{8\pi}{\psi}\frac{\rho_m - 3p_m}{3 + 2\omega}, \label{eq3}
\end{eqnarray}
and the conservation of the matter energy momentum tensor gives
\begin{equation}
\dot{\rho}_m + (\rho_m + p_m)\frac{3\dot{a}}{a} = 0, \label{eq4}
\end{equation}
such that $\rho_m = \rho_{m0}a^{-3}$.
Taking the case $p_{m} = 0$ and substituting (\ref{eq3}) in (\ref{eq2}) and using (\ref{eq1}), we get
\begin{equation}
\frac{H\dot{\psi}}{\psi} - \frac{3\Omega_{m}H^2}{\psi}\left(\frac{1 + \omega}{3 + 2\omega}\right) + \dot{H} + 3 H^2 = 0, \label{eq5}
\end{equation}
where $\Omega_{m} = 8\pi\rho_m/3H^2$.
For the $R_h = ct$ universe, with $a(t) = t/t_{0}$ and $\Omega_m = \Omega_{m0}a^{-1} = \Omega_{m0}H_{0}^{-1}t^{-1}$, Eq. (\ref{eq5}) yields
\begin{equation}
\psi(t) = \frac{3\Omega_{m0}(1 + \omega)}{H_{0}t(3 + 2\omega)}.
\end{equation}
It can easily be checked that this expression for the scalar field satisfies the complete set of field equations in (\ref{eq1})-(\ref{eq4}) provided that the Brans-Dicke parameter $\omega=-2$, which means that $\psi(t) = 3\Omega_{m0}/2H_{0}t$, such that the variable gravitational term $G \sim 1/\psi \sim t$ increases with cosmic time $t$, contrary to Dirac's LNH in which $G\sim1/t$. Hence this, together with fact that the required value of $\omega$ does not satisfy the solar system constraint, imply that this variable $G$ model in BD theory does not offer an appropriate explanation of the dark energy component in the $R_h = ct$ universe. It should be noted that the $R_h = ct$ model $a(t) = t/t_0$, is also a solution of BD-theory with a variable cosmological term, which is represented by using a scalar potential of the type $V(\psi) = 2\psi\Lambda(\psi)$ in (\ref{action}). In this case (see for example \citet{pimentel99}) we get a similar behavior for $G\sim1/t$, while the cosmological term $\Lambda\sim1/t^2$.

\section[]{Quintessence}

In scalar-tensor theory, quintessence, which is represented by a light scalar field $\phi$ which can be minimally, non-minimally or conformally coupled to gravity and having negative pressure, has long been considered (see for example \citet{faraoni00} or \citet{harko14} for recent results) as a possible explanation of cosmic acceleration. \\
The action of scalar-tensor theory is given by
\begin{equation}
S = \int L \sqrt{-g} d^4x, \label{action-st}
\end{equation}
with {Lagrangian density}
\begin{equation}
L= \left(\frac{1}{2\kappa} - \frac{\xi \phi^2}{2}\right)R - \frac{1}{2}g^{ab}\nabla_{a} \phi \nabla_{b} \phi -V(\phi) + L_M,
\label{Lag1}
\end{equation}
where $\kappa=8\pi G = 8\pi$, $\xi$ is the coupling constant between the scalar field and the geometry (with $\xi=0$ representing minimal coupling), $V(\phi)$ is the scalar field potential, and $L_M$ is the {Lagrangian density} associated for the matter distribution.
The variation of the action in (\ref{action-st})
with respect to the scalar field $\phi$ leads to the Klein-Gordon equation
\begin{equation}
\Box\phi - \xi R \phi - \frac{dV}{d\phi} =0, \label{klein-gordon}
\end{equation}
while the variation of the action with respect to the metric $g_{ab}$ gives
\begin{equation}
(1-\phi^2\xi \kappa) (R_{ab}-\frac{1}{2}g_{ab}R)= \kappa (\theta_{ab}+ T_{ab}),
\label{fe1}
\end{equation}
where
\begin{equation}
\theta_{ab}=\nabla_{a} \phi \nabla_{b} \phi - \frac{1}{2} g_{ab} \nabla^c \phi \nabla_c \phi - Vg_{ab}+ \xi ( g_{ab}\Box \phi^2 - \nabla_{a} \nabla_{b} \phi^2),
\label{sems}
\end{equation}
represents the energy momentum tensor corresponding to the scalar field $\phi$ and
\begin{equation}
T_{ab} = (\rho_m + p_m)u_{a}u_{b} + p_m g_{ab}.
\end{equation}
Writing (\ref{fe1}) in the form $G^{a}_{b} = \kappa(\tilde{\theta}^{a}_{b} + T^{a}_{b})$, where $\tilde{\theta}^{a}_{b} = \theta^{a}_{b} + \xi\phi^2G^{a}_{b}$ and using the contracted Bianchi identities $\nabla_{a}G^{a}_{b} = 0$, we get
\begin{equation}
\partial_{b}\phi(\Box\phi - \xi R\phi - \frac{dV}{d\phi}) + \nabla_{a}T^{a}_{b} = 0,
\end{equation}
so that by the Klein-Gordon equation in (\ref{klein-gordon}) we get the separate conservation of the matter energy momentum tensor $T_{ab}$.\\
For the FRW metric in (\ref{frw}) the field equations (\ref{klein-gordon}) and (\ref{fe1})  give for the minimally coupled case ($\xi=0$),
\begin{equation}
H^2  =  \frac{8\pi}{3}(\rho_{m} + \rho_{\phi}),
\end{equation}
\begin{equation}
\dot{H}  +  H^2  =  -\frac{4\pi}{3}(\rho_{m} + \rho_{\phi} + 3p_{\phi}),
\end{equation}
and
\begin{equation}
\ddot{\phi} + 3H\dot{\phi} + V'(\phi) = 0,
\end{equation}
respectively, where for a dust matter distribution $p_{m} = 0$, $\rho_m = \rho_{m0}a^{-3}$ (which follows from the conservation of the matter energy momentum tensor) and $\rho_{\phi} = \dot{\phi}^2/2 + V(\phi)$ and $p_{\phi} = \dot{\phi}^2/2 + V(\phi)$. So for the $R_{h} = ct$ universe with $a(t) = t/t_{0}$, $H(t) = 1/t$ and $\Omega_{m} = \Omega_{m0}a^{-1}$, these equations can be easily solved to give,
\begin{equation}
\phi(t) = \frac{1}{\sqrt{\pi}}\ln[2\sqrt{H_0 t} + \sqrt{2}\sqrt{2 H_0 t - 3\Omega_{m0}}] - \frac{\sqrt{2 H_0 t - 3 \Omega_{m0}}}{\sqrt{2\pi H_0 t}},
\end{equation}
and
\begin{equation}
V(t) = \frac{4 H_0 t - 3 \Omega_{m0}}{16\pi H_0 t^3},
\end{equation}
such that
\begin{equation}
\rho_{\phi} = \frac{3(H_0 t - \Omega_{m0})}{8\pi H_0 t^3}, \qquad p_{\phi} = -\frac{1}{8\pi t^2}.
\end{equation}
The scalar field is well defined and satisfies the weak energy condition $\rho_{\phi} \geq 0 \quad p_{\phi} + \rho_{\phi} \geq 0$ provided that $t \geq 3\Omega_{m0}/2H_0$. The EOS of the dark energy component represented by the scalar field, is given by
\begin{equation}
\omega_{\phi} = p_{\phi}/\rho_{\phi} = \frac{-H_0 t}{3 H_0 t - 3\Omega_{m0}},
\end{equation}
For large $t$ when the scalar field dominates over the matter distribution $\omega_{\phi} \rightarrow -1/3$ in accordance with the EOS $p = -\rho/3$ between the total pressure and total energy density in the $R_h = ct$ universe.

\section{Conclusion}

In this paper we have examined possible sources for the dark energy component of the $R_h = ct$ universe, that are consistent with the total EOS $p = -\rho/3$. We have limited our study to second order theories. It should be pointed out that linear cosmology may also be a solution in higher order theories. For example a vacuum closed FRW model with $a(t) = t/\sqrt{3}$ in the absence of dust is a solution \citep{federbush00} of the gravitational theory with a curvature quadratic action density of the type $(-g)^{1/2}(R_{ik}R^{ik} + bR^2)$, where $b\neq-1/3$.

In our case for simplicity, and also due to the fact that at the present time the radiation component is insignificant, we have assumed that the total pressure $p$ and total energy density $\rho$ are made up of matter and dark energy components. This assumption is also made in most dynamical dark energy models found in the literature that are based on one or more of the sources presented here. In our case we have found that a time dependent cosmological term $\Lambda(t)$ in general relativity or variable gravitational constant $G(t)$ in Brans-Dicke theory are not appropriate sources for the dark energy component in the $R_h = ct$ universe. As pointed put by \citet{basilakos09} the first model has problems with structure formation via gravitational instabilities, while the second model will require $G(t)$ to increase linearly with cosmic time and also requires a Brans-Dicke parameter $\omega = -2$, which is nowhere near the solar system constraint on $\omega$. This will therefore leave quintessence in general relativity as a viable source of the dark energy component in the $R_h = ct$ universe, particularly for $t > 3\Omega_{m0}/2H_0$, where the minimally coupled scalar field (quintessence) is defined and satisfies the weak energy condition. As expected the variable EOS $\omega_{\phi} = p_{\phi}/\rho_{\phi}$ approaches the total EOS $\omega = p/\rho = -1/3$ of the $R_h = ct$ universe for late cosmic time when the scalar field dominates over the matter component. Moreover in this approach the matter and dark energy components are separately conserved as shown in the previous section. This would therefore negate the recent claim made by \cite{lewis13} that the  EOS $R_h=ct$ universe is inconsistent with $p/\rho = -1/3$, or the later claim made by \cite{mitra14} who suggested that the metric in (\ref{frw}) with $a(t) = t/t_0$ is a static vacuum solution. Both claims have already been rebutted by \cite{melia15c} himself, but in this paper we go a step further and obtain a possible explanation of the dark energy component which fits well with the required EOS for the $R_h = ct $ universe.

%

\bsp

\label{lastpage}

\end{document}